# Speciation and dissolution of hydrogen in the proto-lunar disk

Kaveh Pahlevan[1,2] *, Shun-ichiro Karato[2], Bruce Fegley[3]

1. Lagrange Laboratory, Observatoire de la Côte d'Azur, Nice, France.
2. Department of Geology and Geophysics, Yale University, New Haven, CT 06520 USA.
3. Planetary Chemistry Laboratory, Department of Earth and Planetary Sciences and The McDonnell Center for the Space Sciences, Washington University, St. Louis, MO 63130 USA.
   *To whom correspondence should be addressed; Email: pahlevan@oca.eu

**Abstract**
Despite very high temperatures accompanying lunar origin, indigenous water in the form of OH has been unambiguously observed in Apollo samples in recent years. Such observations have prompted questions about the abundance and distribution of lunar hydrogen. Here, we investigate the related question of the origin of lunar H: is the hydrogen observed a remnant of a much larger initial inventory that was inherited from a "wet" Earth but partly depleted during the process of origin, or was primordial hydrogen quantitatively lost from the lunar material, with water being delivered to lunar reservoirs via subsequent impacts after the origins sequence? Motivated by recent results pointing to a limited extent of hydrogen escape from the gravity field of the Earth during lunar origin, we apply a newly developed thermodynamic model of liquid-vapor silicates to the proto-lunar disk to interrogate the behavior of H as a trace element in the energetic aftermath of the giant impact. We find that: (1) pre-existing H-bearing molecules are rapidly dissociated at the temperatures considered (3,100-4,200 K) and vaporized hydrogen predominantly exists as $OH(v)$, $H(v)$ and $MgOH(v)$ for nearly the full range of thermal states encountered in the proto-lunar disk, (2) despite such a diversity in the vapor speciation – which reduces the water fugacity and favors hydrogen exsolution from co-existing liquids – the equilibration of the vapor atmosphere with the disk liquid results in significant dissolution of H into proto-lunar magmas, and (3) equilibrium H isotopic fractionation in this setting is limited to < 10 per mil and the "terrestrial" character of lunar D/H recently inferred should extend to such a precision if liquid-vapor equilibration in the proto-lunar disk is the process that gave rise to lunar hydrogen. Taken together, these results implicate dissolution as the process responsible for establishing lunar H abundances.

## 1. Introduction

After decades of scientific thought maintaining that the Moon is essentially dry (<1ppb $H_2O$), lunar samples have, in recent years, become recognized as carriers of measurable indigenous water (Hauri et al., 2015). In this work, we use "indigenous" to refer to non-solar-wind, non-cosmogenic sources such as inheritance from birth or implantation via impactors (Bottke et al., 2010) and "water" to refer to various forms of hydrogen in silicate glasses, minerals and melts (mostly as OH) with concentrations representing the weight percent of $H_2O$ that would be released upon degassing. In a pioneering study, (Saal et al., 2008) measured tens of ppm wt $H_2O$ in lunar volcanic glasses and inferred much higher $H_2O$ contents in the pre-degassed magma. In a follow-up study, (Hauri et al., 2011) analyzed olivine-hosted melt inclusions from lunar pyroclastic glass deposit 74220 ("orange glass") sampling a pre-eruptive magma and found upwards of hundreds of ppm wt $H_2O$, confirming the earlier inference of high water contents for the source magma. These discoveries have motivated a number of subsequent investigations intended to reevaluate the abundance and distribution of lunar water.

(Karato, 2013) reanalyzed the lunar geophysical observables, exploiting the high sensitivity of electrical conductivity to the presence of protons to infer bulk Moon $H_2O$ abundances of 10-100 ppm wt. (Sharp et al., 2010) observed a wide variation in the chlorine isotopic composition of lunar samples and argued that such variation was due to a dearth of $H_2O$ in the host lava and that the water content of the lunar mantle was constrained to < 0.1 ppm wt. In a subsequent publication, (Sharp et al., 2013) showed that rapid degassing of $H_2$ could dehydrate lunar magmas and reconcile much more hydrous magma compositions with the anhydrous conditions inferred by the chlorine isotopic data. (Hui et al., 2013) measured $H_2O$ contents in plagioclase grains in anorthosite and troctolite from the lunar crust, inferring an initial water content of ~320 ppm wt for the lunar magma ocean from which these grains presumably crystallized. (Albarède et al., 2015) measured Zn, Rb and K abundances in a range of lunar samples, constructed a volatility scale, extrapolated the behavior of moderately volatile elements to H and concluded that the water content of the bulk Moon is ≤ 1 ppm wt. Finally, (Chen et al., 2015) analyzed melt inclusions from a range of mare basalts and found H contents spanning 5-100 times less than that measured by (Hauri et al., 2011), interpreted this range as a signature of degassing and inferred minimum $H_2O$ concentrations for the primitive lunar mantle of 110 ppm wt.

The range in the inferred lunar water abundances attests to the uncertainties in interpreting the lunar volatile record. For sample studies, this uncertainty can arise from: (i) the question of whether the samples in question are characteristic or anomalous in the lunar interior, (ii) an incomplete knowledge of mineral-melt partition coefficients necessary to infer magma compositions from crystalline samples, and (iii) the process of degassing, which, for extrusive volcanism, can eliminate the vast bulk of the inventory of volatiles in magmas, but leave a record from which the history is difficult to reconstruct. Geophysical studies have their own uncertainties; for example, interpretation of the observed electrical conductivity and tidal dissipation in the lunar interior is subject to uncertainty partly due to an incomplete understanding of the sensitivity of the observables to water contents and partly due to an incomplete knowledge of the values of the geophysical observables themselves (Karato, 2013). Such uncertainty allows a variety of inferences to be made from the observable lunar volatile record, with recent inferences ranging from 1 ppm wt $H_2O$ (Albarède et al., 2015) to 110 ppm wt (Chen et al., 2015). For the purposes of this study, we assume that the bulk lunar H abundances are within this range.

Here, we investigate the question of the origin of lunar water: is the measured H a remnant of primordial hydrogen inherited at birth or is it implanted via late-accretionary impacts by, perhaps, water-rich impactors? The discovery of indigenous lunar water has come amidst modern isotopic observations that provide new evidence for a "terrestrial" origin of the lunar material (Kruijer et al., 2015; Pahlevan, 2014; Touboul et al., 2015; Young et al., 2016) motivating new developments to the giant impact hypothesis (Canup, 2012; Cuk and Stewart, 2012; Pahlevan and Stevenson, 2007; Pahlevan et al., 2011) and rendering possible the inheritance of lunar water from the proto-Earth. Motivated by the nearly closed-system behavior of hydrogen in the post-impact Earth-disk system (Nakajima and Stevenson, 2014), we investigate whether the processes accompanying lunar origin can reproduce the bulk elemental abundances recently inferred for the lunar interior. We address this question by constructing a model capable of calculating snapshots of the evolution of the proto-lunar disk and the equilibrium partitioning of hydrogen and its isotopes between the proto-lunar liquid and co-existing vapor to gain insight into the behavior of H in the aftermath of the Moon-forming impact. If the "water" being measured is indeed primordial (i.e. inherited at birth), then its isotopic composition can be used to constrain the process of volatile depletion with the advantage that hydrogen isotopes may exhibit significant equilibrium fractionation even at the high temperatures encountered after the giant impact (see Section 4.2). While a detailed understanding of the processes that established the lunar volatile abundances is – at

present – non-existent, the process of liquid-vapor partitioning illustrated here must play a role in any satisfactory scenario of volatile depletion. In Section 2, we describe the adopted disk model. Section 3 covers the thermodynamic model we use to characterize high-T major element silicates – and the chemical conditions they establish – in the proto-lunar disk, and the modeled behavior of hydrogen as a passive tracer. In Section 4, results of elemental and isotopic partitioning between the silicate vapor atmosphere and an underlying disk liquid layer are presented. In Sections 5 and 6, the implications of these results for the origin of lunar water are discussed and the conclusions of the study are summarized.

## 2. Disk Model

At present, only one evolutionary model of the proto-lunar disk has been developed that attempts to make a connection between a disk process and a chemical or isotopic signature in the lunar material (Canup et al., 2015). Such an evolutionary model is beyond the scope of the present work. Instead, here we describe a physical-chemical model that permits calculation of snapshots of the disk evolution, including the physical and chemical conditions (e.g. $T$, $fO_2$) that determine trace element and isotopic partitioning between the liquid layer and the atmosphere as a function of the thermal state (Figure 1). With this model, we can explore the behavior of tracers for the full range of conditions encountered in the proto-lunar disk during its lifetime.

The physical model is of a 1-dimensional stratified column of disk material with a convective liquid layer near the mid-plane, a liquid-vapor interface, and an overlying convective atmosphere. The partitioning calculations here pertain to the conditions at the liquid-vapor interface, as it is this setting where the composition of the liquid layer and overlying vapor atmosphere may be determined. The two reservoirs (the magma layer and atmosphere) are expected to chemically equilibrate on a short timescale for two reasons: (1) convective turnover timescales are short relative to the lifetime of the proto-lunar disk, which is ~$10^2$-$10^3$ years (Pahlevan and Stevenson, 2007) and (2) on the modern Earth, where $pCO_2$ equilibration between liquid water oceans and the atmosphere occurs on a timescale of ~$10^2$ years (Archer et al., 2009), the exchange is mediated by bubble plumes (Asher et al., 1996). That the proto-lunar magma layer co-exists with an overlying vapor pressure atmosphere implies that the liquid near the interface will be boiling. An argument from analogy with the modern Earth suggests that this feature of the liquid-vapor interface will shorten equilibration timescales further. To calculate equilibrium partitioning, we need an expression for the pressure of equilibration – determined by the atmospheric overburden – as a function of the disk parameters. We assume vertical hydrostatic equilibrium:

$$\frac{1}{\rho}\frac{dp}{dz} = -\frac{GM_E}{r^3}z \quad (1)$$

where $\rho$ is the density of the vapor, $p$ is the pressure, $z$ is the height above the mid-plane, $G$ is the gravitational constant, $M_E$ is the mass of the Earth and $r$ is the cylindrical distance from Earth's spin axis. Following (Ward, 2012), we assume that the vertical component of Earth gravity dominates over the disk self-gravity. To good approximation, the equation of state for the overlying vapor atmosphere can be written:

$$p = \rho k T / \mu \quad (2)$$

where $k$ is Boltzmann's constant, $T$ is the temperature, and $\mu$ is the mean molecular weight of the vapor. Prior works have shown that the two-phase disk is nearly isothermal vertically, i.e. 4 orders of magnitude of pressure variation can be accommodated with a 2-fold change in the temperature (Thompson and Stevenson, 1988). For simplicity, we adopt an isothermal atmospheric vertical structure. Combining (1) and (2) and integrating, we have:

$$\rho(z) = \rho_{int} \exp(-z^2 / H^2) \qquad (3)$$

$$H = \sqrt{2} C_s / \Omega \qquad (4)$$

where $\rho_{int}$ represents the interface density, $H$ is the scale height, $C_S$ is the isothermal sound speed [$=(kT/m)^{1/2}$] and $\Omega$ is the Keplerian angular frequency [$=(GM_E/r^3)^{1/2}$] at disk radius $r$. The surface density of the vapor atmosphere can be expressed as:

$$\sigma_v = \int_{-\infty}^{\infty} \rho(z) dz \qquad (5)$$

For the density structure described in equation (3), integration yields $\sigma_v = \sqrt{\pi} \rho_{int} H$, with $\rho_{int}$ the interface gas density, which, together with equations (2) and (4), yields an expression for the liquid-vapor interface pressure:

$$P_{int} = \frac{1}{\sqrt{2\pi}} f_v \sigma_T C_s \Omega \qquad (6)$$

where $f_V$ is the atmospheric mass fraction of the column (column = atmosphere + liquid layer) and $\sigma_T$ is the total column surface density ($\sigma_v = f_v \sigma_T$). It is this relationship between the equilibration pressure ($P_{int}$) and column vapor fraction ($f_V$) that determines the thermodynamic partitioning described below. This calculation of the pressure – while neglecting the liquid layer – well-approximates the liquid-vapor interface pressure in the relevant limit where the vertical thickness of the liquid layer is much less than the atmospheric scale height, H.

To make a connection with observables, we must relate the calculated composition of the proto-lunar disk liquid and atmosphere to that of the lunar magma ocean. Motivated by the evolutionary model of the proto-lunar disk described by (Ward, 2012), we assume that the disk liquid whose composition is calculated here becomes the lunar magma ocean without further fractionation during the process of lunar accretion (see Section 5.2). In this picture, the liquid layer undergoes "patch" instabilities in which patches of the disk clump via self-gravity before being sheared apart by tidal forces, while the overlying atmosphere remains gravitationally stable. The resulting viscous dissipation near the midplane leads to spreading of the disk liquid beyond the Roche radius where the same process of self-gravity can proceed to generate moonlets (Salmon and Canup, 2012). That the condensation of the lunar material occurs at higher pressures than the triple point of the major silicate components (Nagahara et al., 1994) implicates liquids in the proto-lunar disk, a feature that renders the proto-lunar material more retentive with respect to hydrogen than corresponding solar nebula condensates (Karato, 2013). We select disk parameters ($C_S = 10^3$ m/s, $\Omega = 2.2 \times 10^{-4}$ s$^{-1}$) appropriate to conditions at the Roche radius due to the expectation that the proto-lunar liquid undergo last equilibration with a vapor atmosphere in this setting before fragmenting into moonlets and accreting onto the proto-Moon (Salmon and Canup, 2012).

## 3. Thermodynamic Model

3.1. Major elements

We use a 2-phase binary olivine [(Fe,Mg)$_2$SiO$_4$] thermodynamic model recently developed to study liquid-vapor fractionation in the aftermath of the giant impact (Pahlevan et al., 2011) to capture the physical and chemical conditions prevailing during the thermal evolution of the melt-vapor proto-lunar disk. This choice is motivated by the fact that the Moon is known to be composed primarily of ferromagnesian silicates. In brief, the model describes high-temperature silicate liquids as a binary solution between olivine end-members and silicate vapors as a mixture of ideal gases consisting of the vapor species MgO(v), Mg(v), FeO(v), Fe(v), SiO$_2$(v), SiO(v), Si(v), O$_2$(v) and O(v). Ionized species are known to be present at only trace levels in this environment (Visscher and Fegley, 2013) and are neglected. Input parameters for the model are the total column composition, which we set equal to that of the terrestrial mantle (Fe/Fe+Mg=0.1), the column specific entropy ($S$ = 0.546-1.092 kJ/mol.K, where mol refers to moles of olivine units or silicon atoms) which is varied to capture the range of possible thermal states from fully vaporized to fully condensed, and the pressure of equilibration determined by the overlying atmosphere (Equation 6). Using these relations, we solve for the temperature ($T$), the column vapor fraction ($f_V$), the specific entropy ($S_L$, $S_V$) and composition ($X_L$, $X_V$) of the disk liquid layer and atmosphere, and the partial vapor pressure of each of nine species ($P_i$) for the full range of conditions encountered in the proto-lunar disk. Results from major element calculations include equilibration temperature ($T$) and the oxygen fugacity ($fO_2$) characterizing the liquid-vapor equilibrium that can then be used to calculate the trace element and isotopic partitioning as a function of the thermal state (expressed in terms of, e.g. the entropy or vapor fraction of the column). The main assumption in such an approach is that the liquid layer and overlying atmosphere are in thermodynamic equilibrium. We vary the disk silicate composition (Fe/Fe+Mg) in a plausible range (0.1-0.2) (Khan et al., 2007; Warren, 2005) to test for sensitivity of the results and find that final H abundances in the proto-lunar liquids vary in this range by less than 10 percent. The temperatures of equilibration between the liquid layer and atmosphere that we consider (4,200-3,100 K) span the range from nearly full vaporized ($f_v > 0.99$) to nearly fully condensed ($f_v < 0.01$) states.

3.2. Trace elements – H

With the thermodynamic conditions in the disk – determined by the major elements – calculated, the behavior of hydrogen as a trace element can be described. Experimental work on terrestrial magmas has shown that, in the low-concentration limit, the water fugacity in equilibrium with hydrous silicate melts follows a solubility law of the form $f_{H2O} \propto x_L^2$ (Hamilton et al., 1964) where $f_{H2O}$ is the water fugacity and $x_L$ the mass fraction of water dissolved in the co-existing melt. Such a solubility law is widely interpreted to reflect melt dissolution of water vapor via reaction with oxygen ions in the silicate melt to produce OH groups, a proposition supported by infrared spectroscopic studies of hydrogen speciation in silicate glasses (Stolper, 1982). Because the water abundance in proto-lunar liquids are – by experimental standards – low (< 0.1 wt%) and the temperatures high, we expect hydrogen speciation in proto-lunar liquids to be dominated by OH rather than dissolved molecular $H_2O$ (Stolper, 1989). Recent experimental work has also shown that $H_2$ is likely to be a minor species in melts at the low pressures (< 0.1 GPa) and low total H contents prevailing in proto-lunar liquids, even for reducing proto-lunar magmas (Hirschmann et al., 2012). Finally, while no experimental data exists for the solubility of "water" (as OH) in silicate magmas at the temperatures (3,100-4,200 K) and melt compositions of relevance, existing data indicates slightly retrograde (i.e. decreasing with temperature) solubility, with a relatively weak sensitivity to temperature and melt composition (Burnham and Jahns, 1962; Karsten et al., 1982; Yamashita, 1999). Accordingly, we adopt for the solubility of water in

proto-lunar liquids:

$$P_{H_2O}(bars) = 104 \times \left(\frac{x_L^{H_2O}}{0.01}\right)^2 \qquad (7)$$

as measured in a basaltic melt at ~1370 K (Hamilton et al., 1964). Despite the imprecise match of the melt composition to ultramafic proto-lunar melts and a ~3x difference with temperatures prevailing in the proto-lunar system, we do not explicitly model the dependence of water solubility with available thermodynamic models (Silver and Stolper, 1985). Assuming that the observed insensitivity of solubility to temperature and composition extends to the relevant temperatures and compositions, such variations, while not negligible, represent refinements to the picture here described. As in the analogous terrestrial magma ocean and atmosphere system (Fegley and Schaefer, 2014), experimental determination of water solubility in ultramafic silicate melts at the temperatures of relevance will help to clarify the role of temperature and compositional dependence on equilibrium H dissolution in the proto-lunar disk.

The thermodynamic behavior of H-bearing vapors can be readily articulated. The behavior of H in the vapor phase of the proto-lunar disk is distinctly different from that in vapors in modern magmatic systems in two important respects. First, as noted above, the temperatures are 2-3 times higher in proto-lunar vapors than modern magmatic vapors, and this has the effect of favoring thermal dissociation of molecules. Hence, in addition to the usual $H_2O$-$H_2$ equilibrium of magmatic gases determined by the oxygen fugacity of co-existing melts, we must consider dissociation reactions such as:

$$H_2O(v) = OH(v) + H(v) \qquad (8)$$

$$K_8 = \frac{P(OH) \times P(H)}{P(H_2O)} = \exp(-\Delta G_8 / RT) \qquad (9)$$

An equilibrium speciation reaction and associated equilibrium constant is written for each reaction in Table 1 relating the partial pressures of $H_2O(v)$, $H_2(v)$, $OH(v)$ and $H(v)$.

Indeed, as discussed in the next section, in contrast to magmatic systems, $OH(v)$ and $H(v)$ are important species in the silicate vapor atmosphere of the proto-lunar disk. Secondly, because of the high temperatures, the vapor pressure of major element silicate vapors in this system is significant and, for most of the thermal history of the disk, dominant. Hence, in contrast to magmatic systems, reactions between H-bearing vapor species and major element silicate vapors such as $Mg(v)$ must be considered. In addition to the simple O-H vapor chemistry, we have reactions such as:

$$MgOH(v) = Mg(v) + OH(v) \qquad (10)$$

$$K_{10} = \frac{P(Mg) \times P(OH)}{P(MgOH)} = \exp(-\Delta G_{10} / RT) \qquad (11)$$

with P(Mg) adopted from the major element calculations. We list all H-bearing silicate vapor molecules included in speciation calculations and thermodynamic values used for the Gibbs energy of reactions in Table 1.

With relations for both heterogeneous (Equation 7) and homogeneous (Equation 9 and equivalent for all reactions in Table 1) equilibria for the liquid and vapor, the thermodynamic description of hydrogen as a passive tracer is complete. To calculate the coupled speciation and partitioning of H, we write statements of mass balance between the liquid and vapor:

$$x_H = F_v x_v + (1 - F_v) x_L \tag{12}$$

where $x_H$, $x_V$ and $x_L$ are the total mass fraction of H in the lunar disk, vapor atmosphere, and silicate liquid layer, respectively, and $F_V$ is the column vapor mass fraction. In the calculations presented here, we adopt $x_H = 10^{-5}$-$10^{-3}$ (or equivalently [$H_2O$] = 90-9,000 ppm wt) and discuss the rationale for this parameter choice for proto-lunar disk H abundances ($x_H$) at the end of this section. For the mass fraction of H in the vapor, we have:

$$x_V^H = \frac{\sum_i n_i \times P_i}{\mu \times P} \tag{13}$$

where the summation is over all hydrogen bearing species and with $n_i$ the number of hydrogen atoms contributed by different vapor species with partial pressure $P_i$ and $\mu$ and $P$ representing the mean molecular weight (in amu) and total pressure of the silicate vapor derived from major element calculations. Finally, for the mass fraction of H in the liquid, we have:

$$x_L^H = \frac{2}{18} x_L^{H_2O} \tag{14}$$

where the conversion factor arises because H melt abundances in solubility laws (e.g. Equation 7) are typically expressed as weight percent of equivalent water whereas our statements of mass balance are expressed as mass fractions of hydrogen. With equations representing the solubility law (7), speciation reactions (9), (11) and two others for gas phase equilibria (Table 1), and statements of mass balance in (12)-(14), the system is fully specified. We simultaneously solve for the mole fraction of hydrogen species in the vapor (xH, xH$_2$, xOH, xH$_2$O, xMgOH) as well as the mass fraction of hydrogen in the silicate liquid ($x_L^H$) and silicate vapor ($x_V^H$).

What is the total H (or $H_2O$) abundance ($x_H$) in the proto-lunar disk? While this value is not precisely known, several features of the Earth-Moon system constrain the likely range of disk H abundances. First, the latest isotopic data continue to support a terrestrial provenance of the lunar material (Kruijer et al., 2015; Pahlevan, 2014; Touboul et al., 2015; Young et al., 2016). Hence, we can expect that the proto-lunar disk acquired – at minimum – the H abundances of the proto-Earth mantle at the time of the Moon-forming event. Second, both dynamical and cosmochemical lines of evidence suggest that the Earth's water was largely accreted before lunar origin, i.e. the proto-Earth at the time of the Moon-forming impact was significantly "wet". These lines of evidence are as follows: (a) dynamical: the small value of the lunar inclination suggests that the Moon-forming giant impact was a "late" event in Earth accretion history, resulting in a nearly fully-accreted Earth with a mass likely near 0.99 $M_E$ (Pahlevan and Morbidelli, 2015). Moreover, the Moon-forming impact is a local dynamical event in the inner Solar System and there is no known reason that the nature of Earth-forming impactors would fundamentally change with the occurrence of this event. For this reason, models of planetary accretion consistently find that most of the Earth's water was accreted before the late veneer, i.e. before the Earth had acquired 0.99 $M_E$ (O'Brien et al., 2014; Rubie et al., 2015), (b) cosmochemical: the mass accreted to the Earth after the Moon-forming event (the "late veneer") is constrained by the isotopic composition of the silicate Earth and Moon to be a small fraction of the Earth's mass, likely 0.01 $M_E$ or less (Jacobson et al., 2014). Such a small mass can deliver only a fraction of Earth's inventory of volatiles elements (e.g. that of $^{204}$Pb) and the geochemical coherence of volatile elements suggests that a similar argument can be made for other elements whose terrestrial budgets are not precisely known (e.g. H). The cosmochemical arguments for substantial pre-lunar terrestrial accretion of volatile elements are summarized in (Morbidelli and Wood, 2015). Motivated by the observations indicating that the proto-lunar disk was sourced from the Earth and that the silicate Earth's H budget was largely established prior to the Moon-forming event, we adopt $x_H = 10^{-5} - 10^{-3}$ (i.e. [$H_2O$] = 90-9000 ppm wt), a range that encompasses recent inferences of the H budget of the silicate Earth (Hirschmann, 2006; Marty, 2012).

## 4. Results

4.1. Vapor speciation

The H-vapor speciation in the proto-lunar disk is found to be distinct from other petrologic and nebular systems. Speciation in the O-H system in this environment is dominated by OH(v) and H(v) rather than the more familiar magmatic species $H_2O$(v) and $H_2$(v) for nearly the full range of conditions encountered in the proto-lunar disk (Figure 2).

Both higher temperatures and lower mixing ratios of H in the silicate-rich vapor favor dissociation of $H_2O$(v) and $H_2$(v) relative to modern magmatic systems. The behavior of hydrogen in this system also differs in important respects from that in the solar nebula, where the most abundant equilibrium H-bearing species at 2,000 K are $H_2$ and $H_2O$ (Larimer, 1967; Lord, 1965). Nebular equilibrium is distinct not only because of the lower temperatures but also the distinct abundances. Since the H/Mg ratio of the solar nebula is ~$10^3$, conclusions about nebular speciation of H can be reached without a consideration of Mg-bearing species. In the silicate proto-lunar disk, by contrast, the situation is reversed with Mg/H ~ $10^3$ and Mg-bearing vapor molecules (e.g. MgOH) have the potential to sequester a significant fraction of the trace H abundance in proto-lunar vapors. A similar situation exists for H-bearing Si- and Fe-species (e.g. SiH). Accordingly, in our calculations, we include metal-bearing molecules bonded with H for which thermodynamic data exist (Table 1). We find that MgOH(v) is an important carrier of hydrogen in the proto-lunar disk and more abundant than $H_2O$(v) and $H_2$(v) at even moderate degrees of silicate vaporization ($f_v > 0.1$). As condensation proceeds, the dominant H-bearing species transition from MgOH(v), OH(v) and H(v) to primarily $H_2O$(v) and $H_2$(v) as the silicate vapor atmosphere transitions to a steam atmosphere (Figure 2) analogous to the steam atmosphere of the Earth (Zahnle et al., 1988). The spectrum of vapor species present during the condensation has consequences for both partitioning (section 4.2) and isotopic fractionation (section 4.3).

4.2. Dissolution

How much hydrogen partitions into proto-lunar liquids? The vapor-phase speciation reactions render OH(v), H(v) and MgOH(v) dominant hydrogen reservoirs in the proto-lunar disk and, for a given hydrogen abundance, decrease the water vapor partial pressure and water fugacity. Hence, in considering water dissolution into silicate liquids, such reactions have the influence of shifting the equilibrium towards exsolution. Nevertheless, we find that significant disk hydrogen (2-22% of the total) remains dissolved in proto-lunar liquids (Figure 3).

The high solubility of H in silicate melts translates into significant proto-lunar liquid H abundances despite the high temperatures, high degrees of vaporization and diverse range of

H-bearing vapor species in the proto-lunar disk. For a range of total H contents for the proto-lunar disk of 90-9,000 ppm wt $H_2O$, proto-lunar liquids with $H_2O$ concentrations of several x 1-100 ppm wt arise. While the giant impact is often assumed to result in catastrophic near-complete degassing of the lunar material, these calculations suggest that equilibrium partitioning in a vapor atmosphere can result in significant retention of H in proto-lunar liquids.

4.3. Equilibrium isotopic fractionation

Unlike potassium isotopes, which predominantly exist in atomic form in silicate vapors (Visscher and Fegley, 2013) and are ionically bonded in the liquid, rendering near-zero equilibrium isotopic fractionation (personal communication, Edwin Schauble, 2009), H in this system exists in a variety of bonding environments ranging from OH groups in the melt, to water vapor, the hydroxyl molecule, magnesium hydroxide and molecular and atomic hydrogen in the vapor. Significant differences in the bonding environment of an element between two phases can cause significant equilibrium isotopic fractionation, even at the temperatures encountered after the giant impact (Pahlevan et al., 2011). Such differences in the bonding environment for hydrogen between the liquid and vapor can, therefore, potentially fractionate the H isotopes even at the relevant high temperatures. Hence, if the water being measured as OH in lunar samples is primordial, then its isotopic composition can potentially be used to constrain the processes of lunar volatile depletion.

Unfortunately, isotopic fractionation factors for hydrogen in silicate melts at the relevant temperatures do not – at present – exist. Moreover, existing experimental data for H isotope fractionation in silicates at igneous temperatures (Dobson et al., 1989; Pineau et al., 1998) does not exhibit the usual high-temperature behavior ($lnB=a/T^2$) ensuring no reliable extrapolation to high temperatures at the present time. Nevertheless, equilibrium isotopic fractionation between the vapor species MgOH(v), $H_2O$(v), OH(v), $H_2$(v), and H(v) are calculable and among the highest in nature. We can therefore gain insight into equilibrium behavior of H isotopes in the proto-lunar disk by making the assumption that the bonding environment for OH in silicate liquids can be approximated by that of the OH molecule in the vapor. The problem is then reduced to calculating the isotopic fractionation between the vapor species.

The isotopic fractionation factor ($B$) for $H_2O$(v) and $H_2$(v) as a function of temperature has been extensively studied experimentally and theoretically. We adopt the results of the calculations of (Richet et al., 1977) for these molecules. OH(v) and MgOH(v), by contrast, have received much less attention as isotopic molecules. The vibrational frequencies of these molecules are known from spectroscopic (Chase et al., 1985) and ab initio (Koput et al., 2002) studies, rendering fractionation factors readily calculable. Here, following the standard procedure (Urey, 1947), isotopic fractionation between molecules are calculated by writing the appropriate exchange reactions between isotopic species:

$$D(v) + OH(v) = H(v) + OD(v) \qquad (15)$$

$$K_{15} = \frac{Q(OD)}{Q(OH)} \bigg/ \frac{Q(D)}{Q(H)} \qquad (16)$$

where $Q$'s refer to partition functions whose ratios for isotopically substituted and normal species can be expressed assuming harmonic vibrations:

$$\frac{Q'}{Q} = \prod_i \frac{v'_i}{v_i} \frac{e^{-U'_i/2}}{e^{-U_i/2}} \frac{1-e^{-U_i}}{1-e^{-U'_i}} \qquad (17)$$

where *v* and *v'* are vibrational frequencies for the normal and isotopically substituted molecules, respectively, with adopted values displayed in Table 2.

We follow this procedure for calculating the fractionation factors for both OH(v) and MgOH(v). The main characteristic determining the isotopic composition of vapor species is whether or not it features the high frequency OH stretch, which strongly concentrates deuterium. For this reason, fractionation factors for $H_2O$(v), OH(v) and MgOH(v) are very similar and distinct from those for $H_2$(v) and H(v). H isotopic fractionation factors for the vapor species are plotted in Figure 4.

These calculations demonstrate that at 3,000 K, the OH(v)-H(v) and OH(v)-$H_2$(v) isotopic fractionation is approximately 60 and 30 per mil, respectively, while fractionations between $H_2O$(v), OH(v) and MgOH(v) are relatively minor. Hence, to the extent that $H_2$(v) and H(v) are significant H-bearing species (Figure 2), measurable isotopic fractionation may accompany liquid-vapor equilibration.

To translate such isotopic calculations into statements about Earth-Moon differences, we must convolve fractionation factors with a physical model that yields relative abundances of the vapor species present as well as the temperatures of liquid-vapor equilibration. For hydrogen isotopic mass balance, we have:

$$\delta_L = \delta_{OH} \qquad (18)$$

$$\delta_V = \sum_i x_i \delta_i \qquad (19)$$

$$\delta_C = F_V^H \delta_V + (1 - F_V^H) \delta_L \qquad (20)$$

where $\delta_L$, $\delta_V$, and $\delta_C$ represent the isotopic composition of the liquid, vapor, and total column, respectively, $\delta_i$ and $x_i$ represent the isotopic composition and fraction of vaporized H atoms in a given species, and $F_V^H = F_V \times (x_V^H / x^H)$, expressing the fraction of total hydrogen partitioned into the vapor. Equilibrium isotopic fractionation between a silicate liquid and vapor containing the equilibrium spectrum of species in the proto-lunar disk is displayed in Figure 5.

Equilibrium D/H isotopic fractionation between the proto-lunar liquid and atmosphere, while not negligible, is small (5-10‰). Hence, a "terrestrial" D/H for lunar water (Barnes et al., 2014; Füri et al., 2014; Saal et al., 2013; Tartèse et al., 2013) is consistent with a terrestrial provenance for the Moon-forming material followed by equilibrium dissolution of H in the proto-lunar disk. Despite large equilibrium fractionations between individual species (~60 per mil between OH(v)-H(v) at 3,000 K), the distinct isotopic compositions of atomic and molecular H are diluted by the existence of other less isotopically distinct vapor species such as OH(v), MgOH(v) and $H_2O$(v). Such calculations are based on the assumption that OH(l) and OH(v) are isotopically indistinguishable. A small residual OH(l)-OH(v) fractionation at these temperatures cannot be excluded, and such a fractionation could shift the results here calculated by, perhaps, several tens of per mil (Dobson et al., 1989; Pineau et al., 1998). Despite this uncertainty, and the difficulty of characterizing the silicate Earth isotopic composition with high precision, one prediction of an equilibrium dissolution origin of lunar hydrogen is that the isotopic composition of juvenile lunar water should be equal to the terrestrial D/H value to a precision of tens of per mil.

## 5. Discussion

5.1. Thermal escape from the Earth-Moon system

Here, we have considered liquid-vapor equilibration as a process capable of determining proto-lunar water abundances under the assumption that the post-impact Earth-Moon system behaved as a closed system. How accurate is this assumption? We consider two thermal escape processes: Jeans and hydrodynamic escape. In the case of Jeans escape, the loss is a gas kinetic process with the escaping molecules passing through a low-density exosphere in which outbound high-velocity molecules have a significant probability of leaving the system without further collision. Defining the exosphere as the location where the mean free path of molecules equals the atmospheric scale height, we derive an exospheric number density of ~$10^7$ cm$^{-3}$. Even assuming that all molecules that impinge on the exosphere escape, the flux of hydrogen that can be eliminated through such a process is limited by the short timescales available (~$10^2$-$10^3$ years) and is orders of magnitude too low to significantly influence the lunar H budget. What about hydrodynamic loss in a planetary wind? Hydrodynamic loss from planetary atmospheres is often discussed in relation to the escape parameter ($\lambda$) defined as the ratio of gravitational potential energy ($GM\mu/r$) to the thermal energy ($kT$) of vapor molecules. For large values of the escape parameter (e.g. atomic nitrogen in modern Earth's upper atmosphere), the atmosphere is stable against hydrodynamic escape and adopts instead a hydrostatic state. At sufficiently small values of $\lambda$, escape of a hydrodynamic nature may take place. By adopting $\lambda_{crit}$ = 3 (Volkov et al., 2011) and evaluating the escape parameter at

the Roche radius ($r = 3\ R_E$), the criterion for hydrodynamic escape can be expressed:

$$\lambda_{SVA} = \left(\frac{\mu}{m_p}\right)\left(\frac{2000K}{T}\right) < 3 \qquad (21)$$

Hence, at the temperatures relevant to the silicate vapor atmosphere, a hypothetical, pure light gas, e.g. pure H(v) or $H_2$(v), would be unstable to with respect to hydrodynamic loss while silicate vapors ($\mu$ = 30-40 $m_p$) are gravitationally bound, stable, and hydrostatic. Moreover, no mechanism is known that can separate the light, unbound, H-bearing species from the heavy silicate vapor molecules efficiently on the relevant timescales in this atmosphere: diffusion-limited gravitational separation – while capable of carrying a small flux of hydrogen across the homopause up to escaping levels – is orders of magnitude too slow (Nakajima and Stevenson, 2014). Hence, in the presence of gravitationally bound silicate vapors, the H in the proto-lunar atmosphere is effectively retained via molecular collisions. Toward the end of the thermal history, when silicate vapors have nearly completely condensed, H-speciation at the liquid-vapor interface transitions from OH(v)-H(v)-MgOH(v) to $H_2O$(v)-$H_2$(v). Whether conditions in such a proto-lunar "steam" atmosphere were conducive to hydrodynamic escape depend on: (1) the H/O ratio of the vapor determined by the redox state of the co-existing magma, (2) the thermal structure of the upper atmosphere and associated speciation equilibria, and (3) the presence or absence of other volatile elements (e.g. C) with high enough abundances to influence the mean molecular weight of the residual vapor. A description of the proto-lunar "steam" atmosphere is beyond the scope of this work – it requires further attention. During the silicate vapor dominated phase of the evolution described here, however, the post-impact Earth-Moon system is essentially closed with respect to H loss.

5.2. Fractionation before magma ocean crystallization

Here, we have considered liquid-vapor equilibration in the proto-lunar disk, only one process in the post-impact evolution: both crystal-liquid-vapor fractionation during lunar accretion and hydrodynamic escape from a "naked" hydrous magma ocean (Pahlevan and Karato, 2011) are distinct possibilities for H loss. What assessments can be made about fractionation in these settings? During the lunar accretion process, disk patches with masses < $10^{-2}$ lunar mass are thought to fragment, generating moonlets that feed lunar accretion (Salmon and Canup, 2012). It is thought that disk fragments form when the disk vapor fraction is sufficiently low ($f_v$ < 0.01-0.001) (Thompson and Stevenson, 1988). Given that $f_v$ < 0.01 in our model is attained at super-liquidus temperatures > 3,100 K, we expect the disk fragments to be solid-free upon fragmentation, and accordingly to retain dissolved H (Karato, 2013). Subsequent evolutionary stages may also be considered conducive to volatile loss as the escape parameters for a "naked" magma ocean on the Moon and its preceding moonlets are orders of magnitude lower than that for escape from the Earth-Moon system. Two effects may, nevertheless, hinder fractionation in such settings: first, liquids that fragment to form moonlets have values of entropy such that they co-exist with a silicate vapor at low pressures. Hence, like the silicate vapor atmosphere of the disk, the hydrogen vapors in such settings will be accompanied by silicate vapors, and it is not clear to what extent escape from such moonlets will be *selective* in nature, a requirement for compositional evolution. Secondly, in the absence of a thick overlying atmosphere, exposed liquids with high radiative fluxes on such bodies might develop rapid quenching and foundering of solid lids, significantly hindering degassing of interior liquids. Moreover, the

K/U ratios of lunar reservoirs are uniform, suggesting that the episode of volatile loss had ceased by the onset of lunar magma ocean (LMO) crystallization, otherwise some LMO reservoirs would have recorded a pre-volatile-depletion signature which – at least in the case of potassium – is not observed.

5.3. Kinetic isotope fractionation

We have here only considered equilibrium isotopic fractionation between the liquid and vapor phases. But in this rapidly evolving system – in analogy with terrestrial meteorology – it is possible to imagine departures from equilibrium driving kinetic isotope fractionation with observable consequences. Kinetic isotope fractionation is constrained via potassium isotopes because of the presently attained precision of the measurements (Humayun and Clayton, 1995), the uniformity of lunar reservoirs with respect to K/U precluding significant degassing of K during petrogenesis (Prettyman et al., 2006), and near-zero equilibrium isotopic fractionation between an ionically bonded element in the melt and a monatomic vapor (Visscher and Fegley, 2013). The Earth-Moon similarity in K isotopes can therefore be used to set constraints on the maximum departure from equilibrium characterizing liquid-vapor exchange in the post-impact environment (Pritchard and Stevenson, 2000). H isotopes, however, exhibit a distinct, unique behavior that might make them sensitive recorders of kinetics even in scenarios where the isotopes of other elements exhibit no such sensitivity. In silicate melts, the diffusivity of OH is known to be much less than the diffusivity of molecular $H_2O$ such that even small equilibrium amounts of $H_2O$ (or $H_2$) might dominate the total melt diffusivity of H. The liquid H abundances here described are sufficiently low such that OH dominates the liquid speciation over dissolved water or hydrogen molecules. Importantly, as the H abundance decreases, the melt $OH/H_2O$ ratio increases, and the kinetics of diffusion become increasingly slower from melts into bubbles where equilibration with a vapor phase can take place. The ultimate H abundance of proto-lunar liquids may therefore be determined – not by equilibrium partitioning – but by a progressively decreasing rate of diffusion. While not considered in the equilibrium scenario here described, in such a regime, the kinetics of diffusion for proto-lunar liquids can become the rate-limiting step for hydrogen loss from silicate melts, with significant accompanying D/H fractionation. Such effects must be the focus of future studies.

## 6. Conclusions

As recently as 2006, lunar interior $H_2O$ contents were considered to be < 1 ppb (Taylor et al., 2006). In a pioneering study, (Saal et al., 2008) measured indigenous lunar water in volcanic glasses, instigating reevaluations of the bulk lunar H abundances that currently range from $\leq 1$ ppm wt (Albarède et al., 2015) to $\geq 110$ ppm wt (Chen et al., 2015). Here, we have investigated whether such recently elevated (ppm-level) inferences of H abundances could have been inherited during the processes accompanying lunar origin. At the Roche radius, where the last equilibration between the proto-lunar disk liquid and vapor may have taken place, the temperatures of equilibration between liquids and vapors can range between 3,100-4,200 K. At these conditions, water vapor undergoes thermal dissociation and reacts with silicate vapor to form an equilibrium spectrum of vapor species, the most abundant of which are hydroxyl (OH), atomic hydrogen (H), and magnesium hydroxide (MgOH) with both water vapor and molecular hydrogen constituting minor H-bearing species, except when the silicate vapor atmosphere has undergone nearly complete condensation. Such vapor speciation has the influence of decreasing the water fugacity and promoting exsolution from proto-lunar liquids. Nevertheless, we find that – despite extensive degassing into a vapor atmosphere – a small but significant fraction of hydrogen (equivalent to 0.02-0.22 of the total) is dissolved in proto-lunar liquids that ultimately become the lunar magma ocean. For a range of disk abundances of H (equivalent to 90-9,000 ppm wt $H_2O$), proto-lunar liquids with $H_2O$ concentrations broadly consistent with recent, elevated (ppm-level) inferences on lunar water contents (Albarède et al., 2015; Chen et al., 2015) can be obtained. Moreover, recent works have shown that – in the presence of silicate vapor – hydrogen in the proto-lunar disk is gravitationally bound. Together, these results suggest that the lunar inventory of hydrogen may be terrestrial water inherited by the proto-lunar liquid through equilibrium dissolution. Hence, just as late accretion of comets/asteroids is thought to have made only minor changes to the H budget of the Earth, late accretion onto the Moon may have been a secondary process as the lunar material may have inherited its water from the Earth and retained it throughout the energetic processes of its formation. The scenario here explored suggests that juvenile lunar "water" should have a D/H that is the same as the terrestrial value to within a few tens of per mil. In the future, experimental determination of the solubility of H – as well as H isotope fractionation factors – in silicate magmas at the temperatures and compositions of relevance may permit improved forward modeling of the consequences of origins scenarios for lunar hydrogen.

**Acknowledgements:** This research was carried out in part through a Bateman Fellowship at Yale University and in part through a Henri Poincaré Fellowship at the Observatoire de la Côte d'Azur (OCA) to K.P. The Henri Poincaré Fellowship is funded by the OCA and the City of Nice. S. K. would like to thank NSF for partial support. B. F. was supported by NASA EPSCOR grant NNX13AE52A. We would like to thank Alessandro Morbidelli for insightful discussions and Francis Albarède and an anonymous reviewer for thorough and helpful reviews.

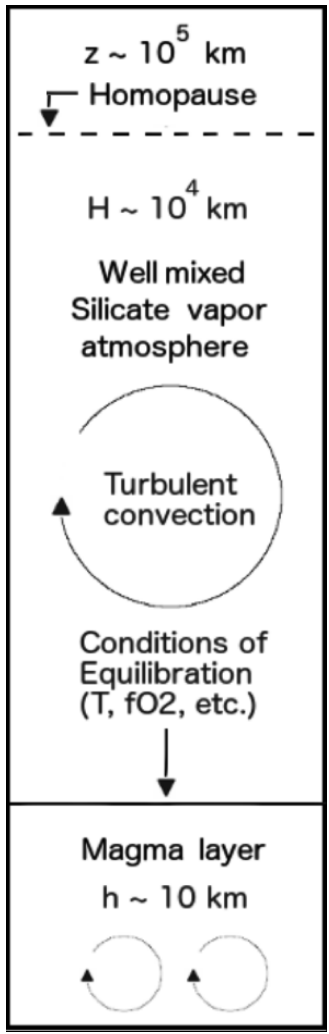

**Fig. 1.** A schematic diagram of a proto-lunar disk column, with a midplane liquid layer, a liquid-vapor interface, and an overlying atmosphere. The physical and chemical conditions at the interface (e.g. T, $fO_2$) determine the partitioning and therefore the composition of the liquid layer and complementary atmosphere. The proto-lunar liquid is assumed to become the lunar magma ocean liquid without further fractionation. The parameter that characterizes the thermal state is the mass fraction of the atmosphere ($f_v$) and can vary between 0-1. Vigorous (convective and other) fluid motions ensure that the vapor atmosphere is largely well-mixed and the homopause – above which vapor species can gravitationally separate according to distinct scale heights – appears only at very high altitudes.

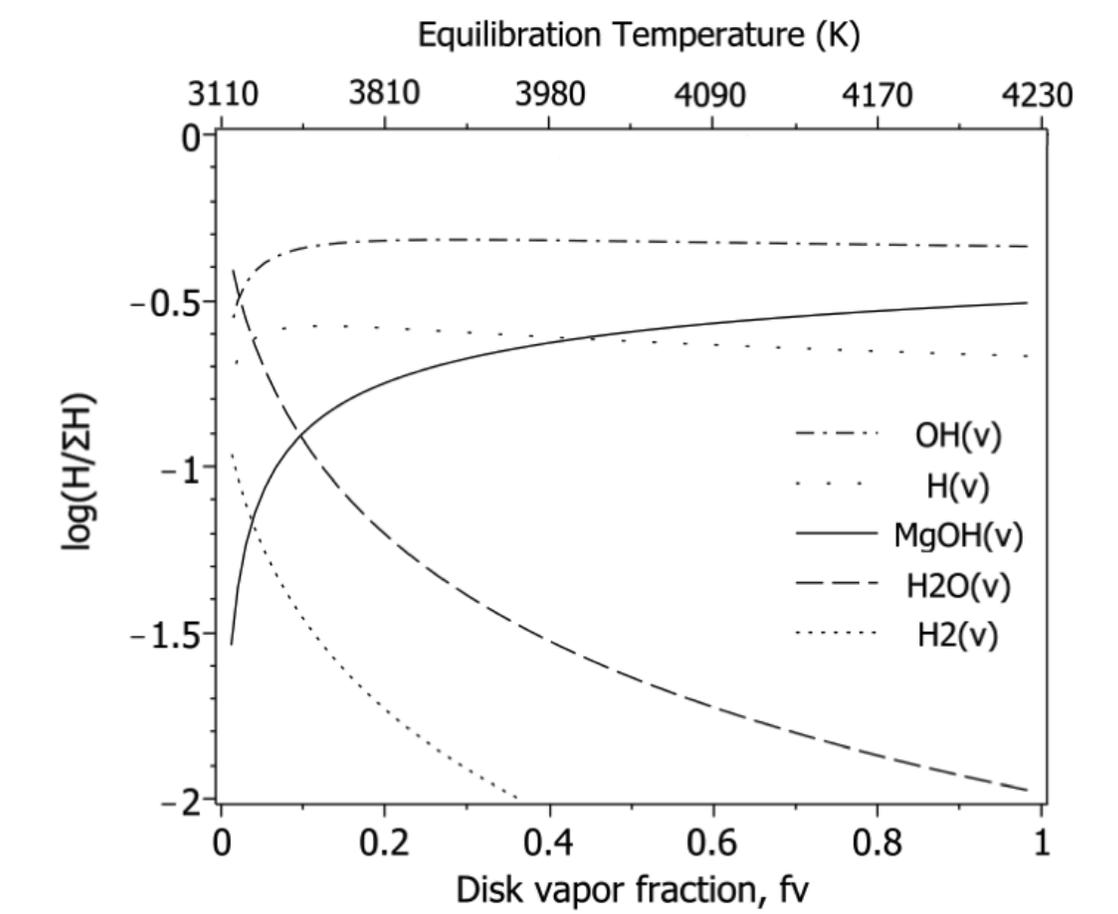

**Fig. 2.** The relative abundances of H-bearing vapor species in the proto-lunar disk as a function of the disk vapor fraction and equilibration temperature for a total disk [$H_2O$] = 900 ppm wt. OH(v) dominates the vapor-phase speciation for the thermal state (*fv*~0.2) generated via the "standard" giant impact (Canup and Asphaug, 2001) as well as more energetic (*fv*~0.5-0.9) high-angular momentum impacts (Canup, 2012; Cuk and Stewart, 2012). Both $H_2$(v) and $H_2O$(v) are trace vapor species until the silicate vapor has undergone near-complete condensation (*fv*~0). The pressure of equilibration is 36 bars at full vaporization ($f_v$=1) and decreases linearly with decreasing vapor fraction (Equation 6).

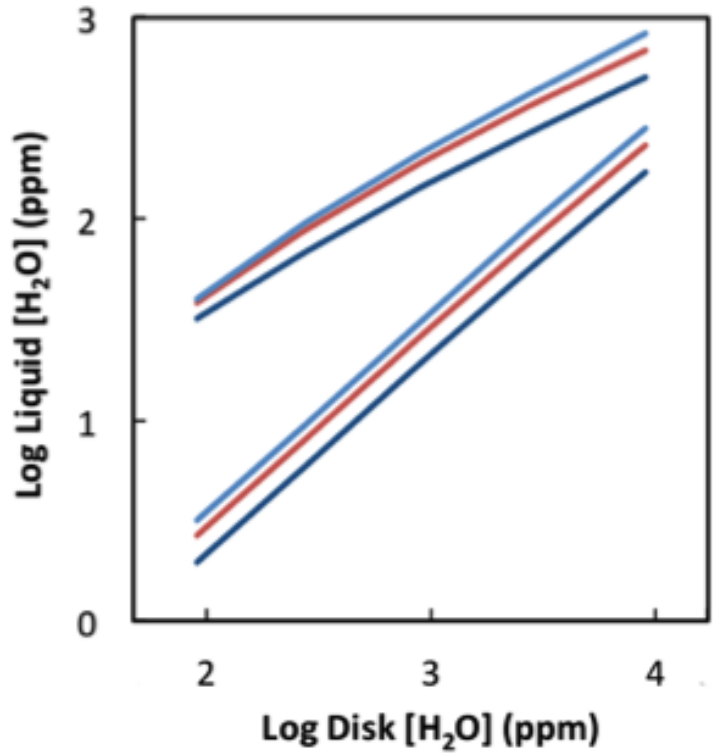

**Fig. 3.** Equilibrium H₂O dissolution into proto-lunar liquids as a function of the total disk H$_2$O abundance. Color denotes total disk surface density of 2 x 10$^7$ kg/m$^2$ (blue), 4 x 10$^7$ kg/m$^2$ (red), and 6 x 10$^7$ kg/m$^2$ (cyan). Higher surface density favors greater dissolution. The lower trio of curves represents the liquid H abundances at the onset of disk condensation ($f_V$ ~ 1) while the upper trio corresponds to conditions toward the end of condensation ($f_V$ ~ 0). The main cause of the offset between the trio of curves is the effect of vapor-phase speciation: predominantly H$_2$O near full condensation, predominantly OH-MgOH-H at higher vapor fractions. OH is expected to dominate the H speciation in the liquid and the presence of other H-bearing liquid species would only make the liquid more retentive with respect to H than here calculated. The liquid H abundances is bounded by pairs of curves, implying significant H-retention into proto-lunar liquids (=0.02-0.22 of the total).

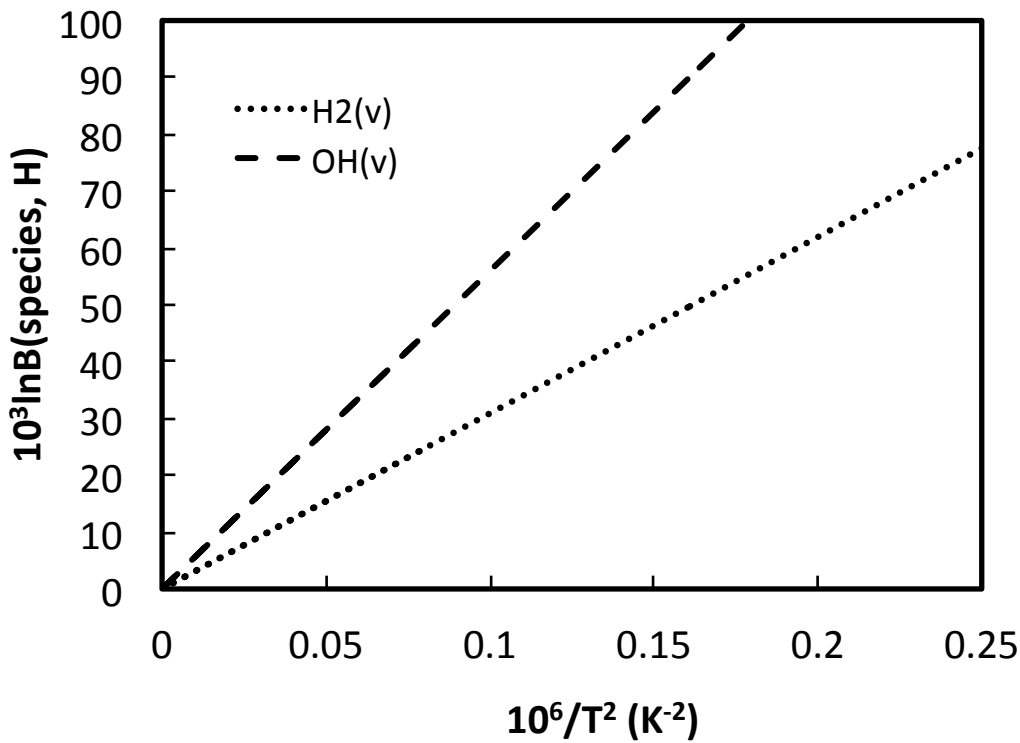

**Figure 4** – Equilibrium isotopic fractionation between co-existing vapor species in the proto-lunar disk. The fractionation factor for $H_2$ is adopted from (Richet et al., 1977). This calculation highlights the importance of H-speciation for articulating the isotopic predictions of liquid-vapor equilibration. Adopted and calculated slopes [=lnB/($10^6/T^2$)] for $H_2O(v)$ and MgOH(v) are 0.59 (Richet et al., 1977) and 0.58, respectively, but given their similarity to that calculated for OH(v) – 0.56 – are not plotted for clarity. The main fractionation in this system is between molecules that do and do not feature the OH stretching mode. The fractionation factor (lnB) for atomic hydrogen is zero by definition.

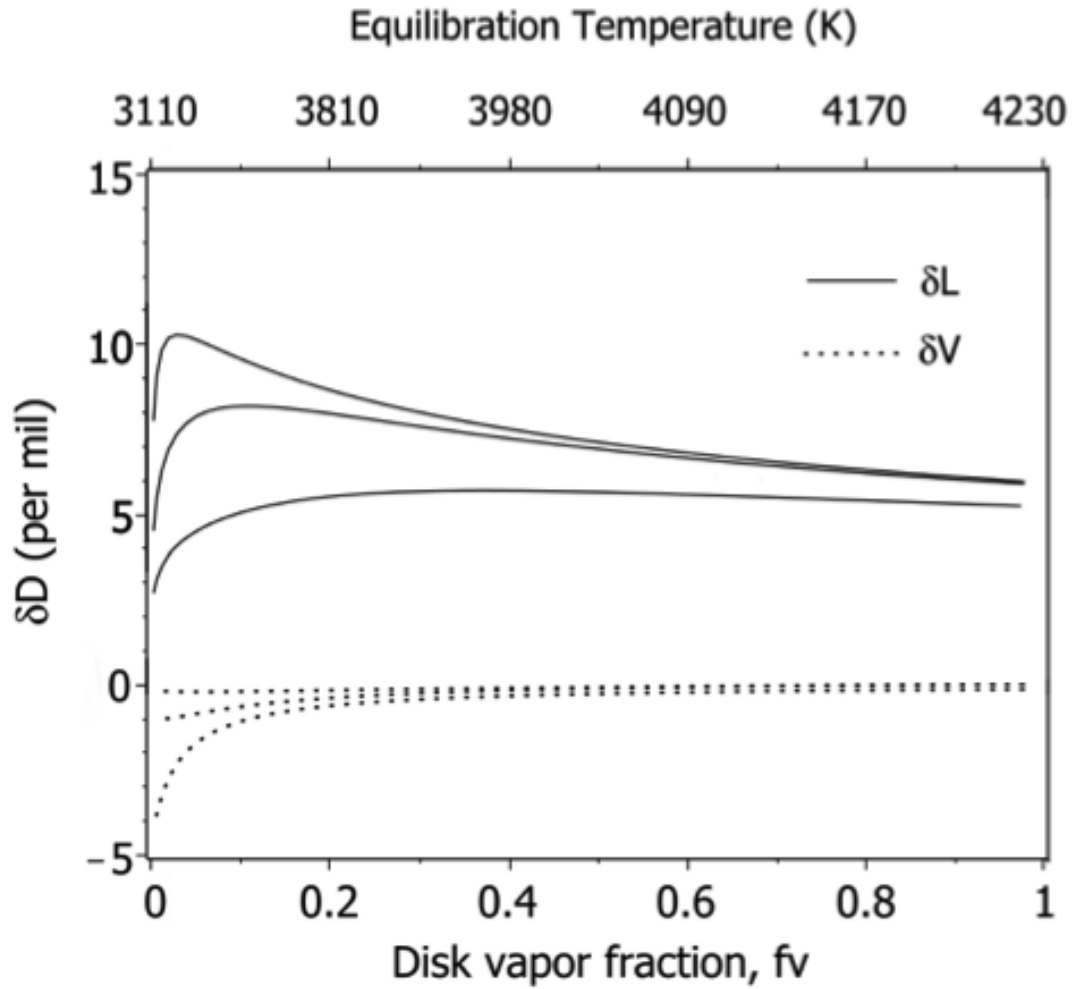

**Fig. 5.** Equilibrium isotopic fractionation between co-existing reservoirs in the proto-lunar disk. The total column (liquid and vapor) composition is taken to be zero ($\delta_C = 0$) and the total disk [$H_2O$] = 90, 900, and 9,000 ppm wt for the three illustrated cases. The equilibrium isotopic fractionation for hydrogen in this 2-phase system is 5-10 per mil. An essentially terrestrial D/H for lunar H is consistent with equilibrium dissolution in the proto-lunar disk. The pressure of equilibration is 36 bars at full vaporization ($f_v=1$) and decreases linearly with decreasing vapor fraction (Equation 6).

**Table 1.** The standard entropy and enthalpy of reactions ($\ln K = \Delta G°/RT$, $\Delta G° = \Delta H° - T\Delta S°$) that determine the vapor speciation of hydrogen in a silicate vapor atmosphere at thermodynamic equilibrium. Thermodynamic data are from (*38*). For the calculations, we consider a number of H-bearing silicate molecules: $Fe(OH)_2$, $Si(OH)_4$, $Si(OH)_2$, $SiOH$, $SiH_4$, $SiH$, $Mg(OH)_2$, $MgOH$ and $MgH$. Only $MgOH(v)$ is present in appreciable abundances in the proto-lunar disk and is therefore explicitly included in the calculations.

| Reaction | $\Delta S°$ (J/mol.K) | $\Delta H°$ (kJ/mol) |
|---|---|---|
| $H_2O(v) = H_2(v) + 1/2\ O_2(v)$ | 58.1 | 250.2 |
| $H_2O(v) = OH(v) + H(v)$ | 132.2 | 514.1 |
| $H_2(v) = 2H(v)$ | 121.4 | 453.6 |
| $MgOH(v) = Mg(v) + OH(v)$ | 104 | 346.6 |

**Table 2** – Vibrational frequencies of isotopically normal and substituted molecules. Fractionation factors derived from these frequencies are plotted in Figure 4.

| Vapor Species | Mode (Degeneracy) | Fundamental Frequency (cm$^{-1}$) | Substituted Fundamental Frequency (cm$^{-1}$) | Reference |
|---|---|---|---|---|
| OH | Stretch (1) | 3735 | 2718 | (Chase et al., 1985) |
| MgOH | Mg-O Stretch (1) | 750 | 737 | (Bunker et al., 1995) |
|  | Bending mode (2) | 161 | 109 |  |
|  | O-H Stretch (1) | 3851 | 2846 | (Koput et al., 2002) |